\documentclass[aps,prb,twocolumn,floatfix,showpacs,citeautoscript,longbibliography,hyperlinks,superscriptaddress]{revtex4-2}

\usepackage{graphicx}
\usepackage[colorlinks=true,citecolor=blue]{hyperref}
\apptocmd{\sloppy}{\hbadness 10000\relax}{}{}
\usepackage{xcolor}
\usepackage{amsmath}
\usepackage[T1]{fontenc}

\newcommand{\revision}[1]{{{#1}}}

\begin{document}
\title{Fast, accurate, and local temperature control using qubits}

\author{Riya Baruah}
\affiliation{Department of Applied Physics, Aalto University, 00076 Aalto, Finland}
\author{Pedro Portugal}
\affiliation{Department of Applied Physics, Aalto University, 00076 Aalto, Finland}
\author{Joachim Wabnig}
\affiliation{Nokia Bell Labs, Cambridge, United Kingdom}
\author{Christian Flindt}
\affiliation{Department of Applied Physics, Aalto University, 00076 Aalto, Finland}
\affiliation{RIKEN Center for Quantum Computing, Wakoshi, Saitama 351-0198, Japan}

\begin{abstract}
Many quantum technologies, including quantum computers, quantum heat engines, and quantum sensors, rely on operating conditions in the subkelvin regime. It is therefore desirable to develop practical tools and methods for the precise control of the temperature in nanoscale quantum systems. Here, we present a proposal for fast, accurate, and local temperature control using qubits, which regulate the flow of heat between a quantum system and its thermal environment. The qubits are kept in a thermal state with a temperature that is controlled in an interplay between work done on the qubits by changing their energy splittings and the flow of heat between the qubits and the environment. Using only a few qubits, it is possible to control the thermal environment of another quantum system, which can be heated or cooled by the qubits.  As an example, we show how a quantum system at subkelvin temperatures can be significantly and accurately cooled on a  nanosecond timescale. Our proposal can potentially be realized with superconducting flux qubits, charge qubits, or spin qubits, which can now be fabricated and manipulated with exquisite control. 
\end{abstract}

\maketitle

\section{Introduction} 

The precise control of temperature at the nanoscale is important for many quantum technologies such as quantum information processing, quantum communication, and quantum sensing~\cite{Zagoskin:2012,Majidi2024}. Thermal fluctuations can introduce unwanted noise, decoherence, and dissipation, which may degrade the functionality of a quantum device. In particular, quantum states are highly sensitive to their thermal environment, which may cause dephasing, loss of coherence and entanglement, and a reduction in the fidelity and efficiency of desired quantum operations. For these reasons, temperature control is needed to mitigate such effects and to extend relaxation and coherence times. Efficient strategies for thermal management may also be required to control the flow of heat in a quantum device~\cite{Pekola2021}. Here, local control of the thermal environment may be used to generate temperature gradients that can guide heat from one part of a quantum circuit to another, where it may be converted into useful work~\cite{Vinjanampathy:2016}.

\begin{figure}
    \centering
\includegraphics[width=0.95\columnwidth]{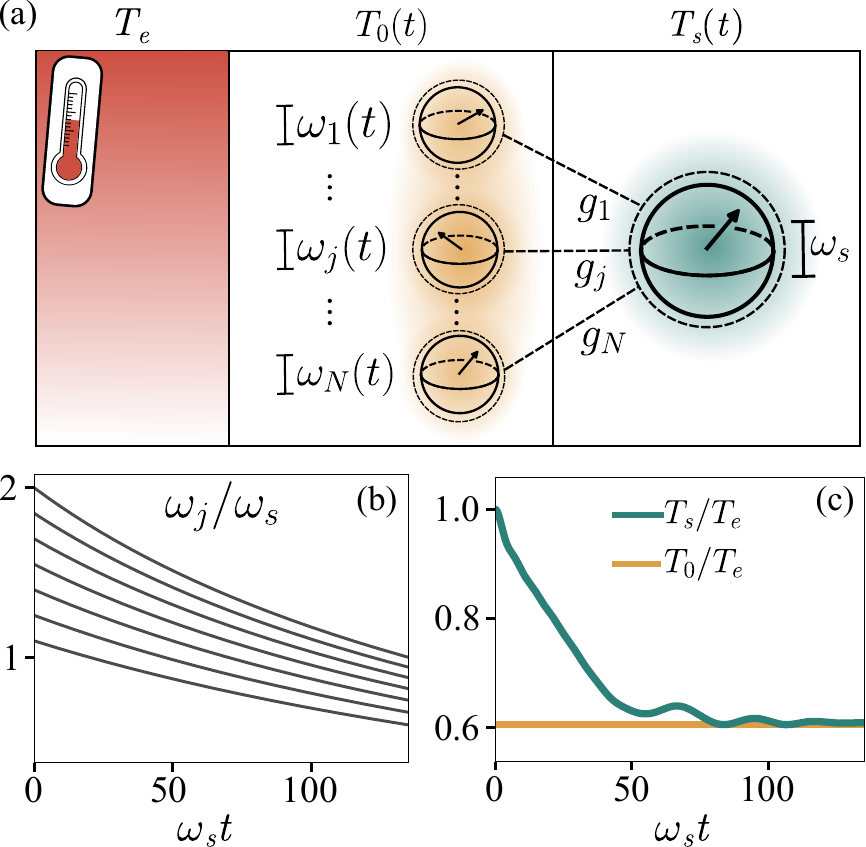}
    \caption{Fast, accurate, and local temperature control. (a) A quantum system, here a qubit (green), is coupled to $N$ other qubits (orange) with the couplings $g_j$, $j=1,\ldots, N$. The energy splittings of the $N$ qubits, $\omega_j$, are modulated in time to control their temperature, $T_0(t)$, which may be different from the temperature of the external reservoir (red), $T_e$. (b) Time-dependent energy splittings for $N=7$ qubits to maintain a constant temperature \revision{$T_0=0.6T_e$} below that of the reservoir. (c) Temperature of the $N=7$ qubits (yellow) and that of the quantum system (green), which is cooled by the qubits.}
    \label{fig:system}
\end{figure}

There are several approaches to temperature control at the nanoscale~\cite{Enss2005,Cangemi:2024}. One common method involves passive cooling, where quantum systems are thermally coupled to cryogenic environments using materials with high thermal conductivity, such as metals, to dissipate heat~\cite{Enss2005}. Electronic refrigeration can be achieved with \revision{NIS} (normal-metal/insulator/superconductor) junctions, where the energy-dependent tunneling of charges selectively removes hot electrons~\cite{Giazotto:2006}. Quantum systems may also be cooled by absorption refrigerators~\cite{Scovil:1959}, which can be realized in different physical setups based on superconductors~\cite{Chen2012,Hofer:2016}, optomechanical resonators~\cite{Mari:2012}, Coulomb-blockade structures \cite{Venturelli:2013,Erdman:2018}, or light-matter interfaces~\cite{Mitchison:2016}. Passive cooling is effective for maintaining a low temperature; however, these methods often lack precision and speed. Active cooling techniques, by contrast, offer dynamic approaches to temperature control. Adiabatic demagnetization, for example, exploits the magnetic properties of materials to cool them through a controlled reduction of an applied magnetic field~\cite{Giauque:1933,Enss2005}. Also,  in a quantum computer, a target qubit can be dynamically cooled through global unitary operations, however, at the cost of heating up other qubits~\cite{Schulman:2005,Brassard:2014,Allahverdyan:2011,Taranto:2023,Bassman-Oftelie:2024}. \revision{In a recent experiment, two auxiliary qudits were used to cool down a target qubit close to its ground state~\cite{Aamir:2025}. Overviews of different temperature control and thermometry techniques can be found in Refs.~\cite{Cangemi:2024,Giazotto:2006,Mehboudi:2019,Jones:2020}.}

Generally, the temperature of a system can be changed in two ways. As an example, a confined gas can be put into contact with a thermal reservoir. If the two have different temperatures, heat will flow between them, and the temperature of the gas will change. This process is irreversible, and entropy is generated.  However, the temperature may also change because work is performed on the gas, for instance, in an adiabatic compression, where the volume is reduced by a piston, which performs work against the pressure~\cite{Pathria2011}. If the compression is fast, there is no time for heat to escape, and the process is adiabatic in the thermodynamic sense. Thus, the work adds to the energy of the gas, whose temperature increases. The process is reversible, and there is no entropy change. 

In this work, we present a proposal for fast, accurate, and local control of the temperature in nanoscale devices using quantum mechanical two-level systems~\cite{Arrachea:2023}. We will refer to the two-level systems as qubits, although they do not necessarily have to be part of a quantum computer. Still, our proposal is motivated by the remarkable progress in fabricating and controlling qubits for quantum information processing~\cite{Krantz2019,Anferov2024,Makhlin:2001,Hanson2007,Burkard:2023}. In particular, it is key to our proposal that the qubit splittings can be accurately controlled in time, which directly translates into an accurate control of temperature as we will see. Specifically, by changing the energy splittings of the qubits and thereby performing work on them, it is possible to realize a finite-size thermal environment~\cite{Mazzola2009,Vega2015}, whose temperature can be controlled. Moreover, in light of recent experimental advances, our proposal appears to be feasible and realistic, and it can be implemented in a variety of physical setups, where accurate, fast, and local temperature control is needed.

Figure~\ref{fig:system} illustrates the idea of the proposal that we  describe in this article. Figure~\ref{fig:system}(a) shows several qubits that function as intermediates between a quantum system (here another qubit) and a thermal reservoir. Initially, the qubits are in equilibrium with the reservoir and have a density matrix that is given by the reservoir temperature. We then extract energy from the qubits by abruptly reducing their energy splittings so that the qubits remain in a thermal state but with a lower temperature. Heat now starts to flow from the reservoir into the qubits because of the temperature difference between them. Next, as depicted in Fig.~\ref{fig:system}(b), to compensate for this heating effect, we further reduce the qubit splittings to maintain the constant temperature indicated in Fig.~\ref{fig:system}(c). Heat then starts to flow from the hotter quantum system into the colder qubits, and the quantum system is cooled down as seen in Fig.~\ref{fig:system}(c). As we will see, it is possible to generate any time-dependent temperature, so that the quantum system for example can be exposed to a periodic temperature drive. Moreover, since our  proposal is rather general, it can be realized with many types of qubits such as superconducting flux qubits~\cite{Krantz2019,Anferov2024}, charge qubits~\cite{Makhlin:2001}, or spin qubits~\cite{Hanson2007,Burkard:2023}. \revision{In the following sections, we describe our proposal in further detail.}

The rest of the paper is organized as follows. In Sec.~\ref{sec:time-dep-temp}, we discuss time-dependent temperatures, and we consider a single qubit that is coupled to a heat reservoir. We illustrate how the temperature of the qubit can be changed by modulating its energy splitting, while heat flows to and from the reservoir. \revision{Later on, we will use several of such qubits to heat or cool a quantum system, which we do not control, as illustrated in Fig.~\ref{fig:system}(a).} In Sec.~\ref{sec:drive_protocol}, we demonstrate how to compensate for the heat flow and drive the qubit so that a desired time-dependent temperature is realized. In Sec.~\ref{sec:eff_temperatures}, we now show how several qubits can be used to fast and accurately control the local thermal environment of another quantum system, for example, by cooling it \revision{down close to its ground state. For illustrative purposes, we take the quantum system to be another qubit, but our proposal can also be applied to other types of quantum systems.} In Sec.~\ref{sec:exp_persp}, we discuss the perspectives of realizing our proposal experimentally using spin qubits as an example. In Sec.~\ref{sec:conclusion}, we conclude our work and provide an outlook on possible avenues for further developments. 

Several technical details of our work are deferred to appendixes. In Appendix~\ref{appA}, we derive the quantum master equation that we use following standard procedures; however, we include the derivation here to illustrate how it is adapted to our time-dependent situation. In Appendix 	~\ref{appC}, we describe the numerical scheme that we use to solve the master equation and time-evolve the driven qubits.

\section{Time-dependent temperatures}
\label{sec:time-dep-temp}

We start by considering a generic quantum system, whose total energy can be modulated in time. Later in this section, we will consider the specific example of a single qubit. To begin with, the system is in equilibrium with a  heat reservoir, such that its density matrix reads 
\begin{equation}
    \hat\rho(0) = \frac{1}{Z_0}e^{-\hat H_0/k_B T_0(0)},
    \label{eq:thermalstate}
\end{equation}
where $Z_0=\mathrm{tr}\{e^{-\hat H_0/k_B T_0(0)}\}$ is the canonical partition function, the Hamiltonian of the undriven system is denoted by $\hat H_0$, and the initial temperature of the quantum system, $T_0(0)=T_e$, is given by that of the reservoir, $T_e$. 

Next, we let the system evolve with a time-dependent Hamiltonian of the form \revision{
\begin{equation}
\hat H_0(t) = w(t)\hat H_0,
\label{eq:sysHam}
\end{equation}
where} the dimensionless prefactor $w(t)$ multiplies the total energy of the undriven system. Such a Hamiltonian can be realized for a single qubit by modulating its energy splitting. However, a similar reasoning would apply to any other quantum system, whose energy can be modulated in time, for example, a quantum harmonic oscillator with a controllable oscillation frequency~\cite{Schumacher:2010,Portugal2022,aNote2}. We first modulate the energy much faster than the timescale over which the system interacts with the heat reservoir, such that the coupling to the reservoir can be ignored. (Below, we return to the case where the coupling to the reservoir is finite.) Considering then only the unitary dynamics, the time-evolved density matrix becomes
\begin{equation}
\hat \rho(t) = \hat U(t,0) \hat \rho(0) \hat U^\dagger(t,0) =  \hat \rho(0),
\end{equation}
where the time-evolution operator $\hat U (t,0)=\hat{T}\{e^{-i\int_0^t dt'
\hat H_0(t')/\hbar}\}$ is given by the time-ordering operator $\hat T$ and the Hamiltonian $\hat H_0(t)$. Importantly, in the last step, we have used that both the time-evolution operator and the initial density matrix are given only by the time-independent Hamiltonian $\hat H_0$ times prefactors, such that they commute. As a result, the density matrix does not change with time. \revision{Now, although the density matrix does not change, the Hamiltonian of the system has changed. Therefore, we  express  the density matrix in terms of the  Hamiltonian at the given time as}
\begin{equation}
    \hat \rho(t) = \frac{1}{Z_0}e^{-\hat H_0(t)/k_B T_0(t)},
    \label{eq:equi_den}
\end{equation}
where we have introduced the parameter 
\begin{equation}
T_0(t) = w(t) T_0(0),
\label{eq:time-dep-temp}
\end{equation}
which we will refer to as the temperature of the system. Indeed, the system would have the density matrix in Eq.~(\ref{eq:equi_den}), if it were in equilibrium with a weakly coupled reservoir at the temperature $T_0(t)$. That aspect will be important in the following. However, we are aware that the definition of temperature is still debated~\cite{Puglisi:2017}, and some might refer to the parameter $T_0(t)$ as an effective temperature~\cite{Quan:2007,Cangemi:2024}. However, for the sake of brevity, we will refer to it simply as the temperature of the system. Moreover, we consider $w(t)$ to be positive, so that the system temperature is always positive.

\begin{figure*}
    \centering
\includegraphics[width=0.95\textwidth]{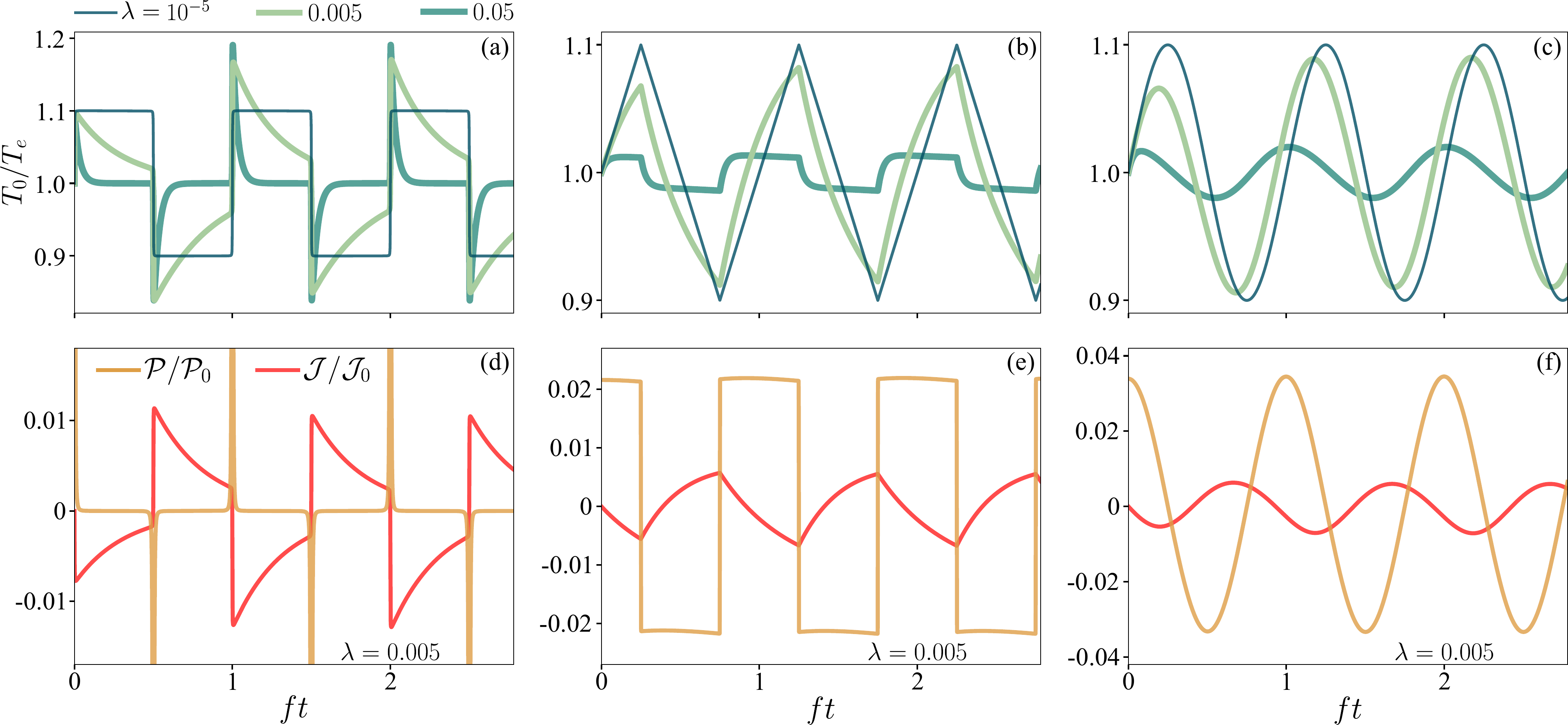}
    \caption{Thermodynamics of a driven qubit. (a) Time-dependent temperature of a qubit that is driven by a periodic square-wave modulation of its energy splitting. We show results for three different couplings to a reservoir at the temperature $T_e$. The driving frequency is $f = 0.1\omega_0$, where $\omega_0=\omega_0(0)$ here is the initial qubit splitting. For the weakest coupling, the temperature directly follows the qubit drive, since $T_0(t)/T_0(0)=\omega_0(t)/\omega_0(0)$ according to Eq.~(\ref{eq:time-dep-temp}). (b),(c) Similar results for a saw tooth drive and a cosine drive. (d)-(f) Power and heat added to the qubit for the coupling $\lambda=0.005$ with $\mathcal P_0= \mathcal J_0 = f\hbar\omega_0$.}  
    \label{fig:1qubit}
\end{figure*}

The temperature of the system changes because we perform work on it. Specifically, the energy measured with respect to the ground state energy $E_0(t)$ evolves as 
\begin{equation}
 U(t) =\mathrm{tr}\{[\hat H_0(t)-E_0(t)]\hat\rho(t)\}=w(t) U(0),
\label{eq:adiab_energy}
\end{equation}
where $U(0)$ is the initial energy. Physically, this situation resembles the adiabatic compression of a gas~\cite{Pathria2011} or cooling of spins by adiabatic demagnetization~\cite{Giauque:1933,Enss2005}. 

While the considerations above are rather general, we now focus on a qubit that is coupled to a heat reservoir. The Hamiltonian of the qubit and the reservoir reads
\begin{equation}
\hat{H}(t)=\hat H_0(t)+\hat H_B+\hat H_I
\end{equation}
with the time-dependent Hamiltonian of the qubit,
\begin{equation}
\hat H_0(t)=\frac{\hbar\omega_0(t)}{2} \hat \sigma_z,
\label{eq:H_qubit}
\end{equation}
\revision{which then takes the form of the Hamiltonian in Eq.~(\ref{eq:sysHam}), and}
 the reservoir is described as harmonic oscillators, 
 \begin{equation}
     \hat H_B = \sum_l \hbar \omega_l \hat a_l^\dagger \hat a_l.
 \end{equation}
Here, the usual bosonic creation and annihilation operators are denoted by $\hat a_l^{\dagger}$ and $\hat a_l$, and the oscillators have the frequencies~$\omega_l$. We have also made use of Pauli matrices to describe the qubit. The interactions between the qubit and the oscillators are given by the Hamiltonian \cite{Haase2018},
\begin{equation}
    \hat H_I = \frac{\hbar}{2} \hat \sigma_x \otimes \sum_l(\kappa_l \hat a_l + \kappa_l^* \hat a_l^\dagger),
    \label{eq:H_I}
\end{equation}
where $\kappa_l$ are the coupling strengths.  With a weak coupling to the reservoir, we can derive a quantum master equation for the density matrix of the qubit, 
\begin{equation}
        \frac{d \hat \rho(t)}{dt} = -\frac{i}{\hbar}[\hat H_0(t), \hat \rho(t)] +  \gamma(\omega_0(t)) \mathcal{D}\hat \rho (t),
    \label{eq:master_single_qubit}
\end{equation}
which consists of two terms. The first term corresponds to the unitary evolution due to the time-dependent Hamiltonian, while the second term describes the exchange of 
energy with the reservoir and reads
\begin{equation}
\begin{split}
    \mathcal{D}\hat \rho =\Big(1 & + n(\omega_0)\Big)\Big(\hat \sigma_- \hat \rho \hat \sigma_+ - \{\hat \sigma_+ \hat \sigma_-, \hat \rho \}/2\Big) \\
    & + n(\omega_0)\Big(\hat \sigma_+ \hat \rho\hat\sigma_- - \{\hat \sigma_- \hat \sigma_+, \hat \rho \}/2\Big),
\end{split}
\label{eq:dissipator}
\end{equation}
where we have left out the explicit time-dependence of the density matrix and the qubit splitting. The dissipator~$\mathcal{D}$ is expressed in terms of the ladder operators $\hat\sigma_\pm = (\hat\sigma_x \pm i \hat\sigma_y)/2$ and the Bose-Einstein distribution   
\begin{equation}
n(\omega_0) = \frac{1}{e^{\hbar \omega_0/k_B T_e} - 1},
\label{eq:BEdist}
\end{equation}
where $T_e$ is the temperature of the environment. We have also defined the bare rate at which heat is exchanged between the qubit and the environment
\begin{equation}
\gamma(\omega_0) =  \lambda \pi\omega_0e^{-\omega_0/\omega_c},
\end{equation}
which depends on the qubit splitting via the spectral function of the reservoir, which we take to be ohmic, although that is not essential for the following discussion. The prefactor $\lambda$ is related to the coupling strengths in Eq.~(\ref{eq:H_I}), and the cutoff frequency $\omega_c$ is taken so high that the exponential function can be replaced by one.

The quantum master equation follows as a special case of the more general derivation in Appendix~\ref{appA}. An important simplification occurs because the qubit Hamiltonian in Eq.~(\ref{eq:H_qubit}) commutes with itself at different times. For that reason, we can essentially proceed as in standard derivations of quantum master equations, and the timescale on which we modulate the qubit Hamiltonian only has to be slower than the fast correlation time of the reservoir.  Moreover, if the qubit initially is in a thermal state as in Eq.~(\ref{eq:equi_den}), it will remain in a thermal state, however, its temperature may change. Thus, we may monitor how the temperature of the qubit changes in response to the time-dependent modulations of the qubit splitting. 

Figures~\ref{fig:1qubit}(a), \ref{fig:1qubit}(b), and \ref{fig:1qubit}(c) show the qubit temperature for a square-wave drive of the qubit splitting, a saw-tooth drive, and a cosine drive, respectively. In each case, we show the qubit  temperature for three different couplings to the reservoir. For the smallest coupling, the qubit temperature follows the time-dependent drive, as anticipated by Eq.~(\ref{eq:time-dep-temp}). As such, this curve also illustrates the time-dependent modulations of the qubit splitting. By contrast, as the coupling is increased, heat starts to flow between the qubit and the reservoir because of the temperature difference between them. We note that the figure shows the transient behavior of the temperature from the onset of the drive, before it becomes periodic.

Surprisingly perhaps, only in Fig.~\ref{fig:1qubit}(a), the qubit  reaches higher (and lower) temperatures with a finite coupling to the reservoir than without the coupling. This observation can be understood as follows. Every time the qubit splitting is increased (or decreased), the temperature increases by the relative change of the qubit splitting, as seen in the case without the coupling. However, with a finite coupling, the qubit temperature has already started to increase toward the temperature of the reservoir. Thus, the temperature is higher at the point, when the qubit splitting is increased, leading to a higher temperature as compared to the uncoupled case.

To understand the thermodynamics of the temperature modulations, we consider the change of the qubit energy, 
\begin{equation}
\begin{split}
     \partial_t U(t) &= \text{tr} \{ \hat  \rho (t) \partial_t [\hat H_0(t)-E_0(t)]\}\\
     &+ \text{tr} \{\hat H_0(t) [\partial_t \hat \rho(t)]\},
\end{split}
\end{equation}
having used that $ \text{tr} \{\partial_t \hat \rho(t)\}=0$ due to trace conservation. From this expression, we can identify the power added to the qubit by changing its energy splitting as
\begin{equation}
     \mathcal{P}(t) = \text{tr} \{ \hat  \rho (t) \partial_t [\hat H_0(t)-E_0(t)]\},
\end{equation}
while the heat current that flows into the qubit reads
\begin{equation}
    \mathcal{J}(t) = \text{tr} \{\hat H_0(t) [\partial_t \hat \rho(t)]\}. 
\end{equation}
Using that the qubit is in a thermal state,  we find
\begin{equation}
     \mathcal{P}(t) =  f(\omega_0(t))\hbar \partial_t\omega_0(t) ,
     \label{eq:q_pow}
\end{equation}
and
\begin{equation}
\begin{split}
    \mathcal{J}(t) = \hbar \omega_0 \gamma(\omega_0) \Big[ & n(\omega_0)(1 - f(\omega_0)) \\  -& (1 + n(\omega_0))f(\omega_0)\Big],
\end{split}
\label{eq:q_heat}
\end{equation}
where we have left out the explicit time-dependence of the qubit splitting in the expression for the heat current. Here, we have used that the occupation of the excited state of the qubit is given by the Fermi function
\begin{equation}
f(\omega_0) = \frac{1}{e^{\hbar \omega_0/k_B T_0} + 1},
\end{equation}
which, in contrast to the Bose-Einstein distribution in Eq.~(\ref{eq:BEdist}), depends on the qubit temperature and not the temperature of the environment. If the temperatures of the two are different, a heat current will flow between them. On the other hand, if they are in thermal equilibrium, the heat current vanishes, which follows from Eq.~(\ref{eq:q_heat}) by setting the two temperatures equal.

\begin{figure*}
    \centering
\includegraphics[width=0.95\textwidth]{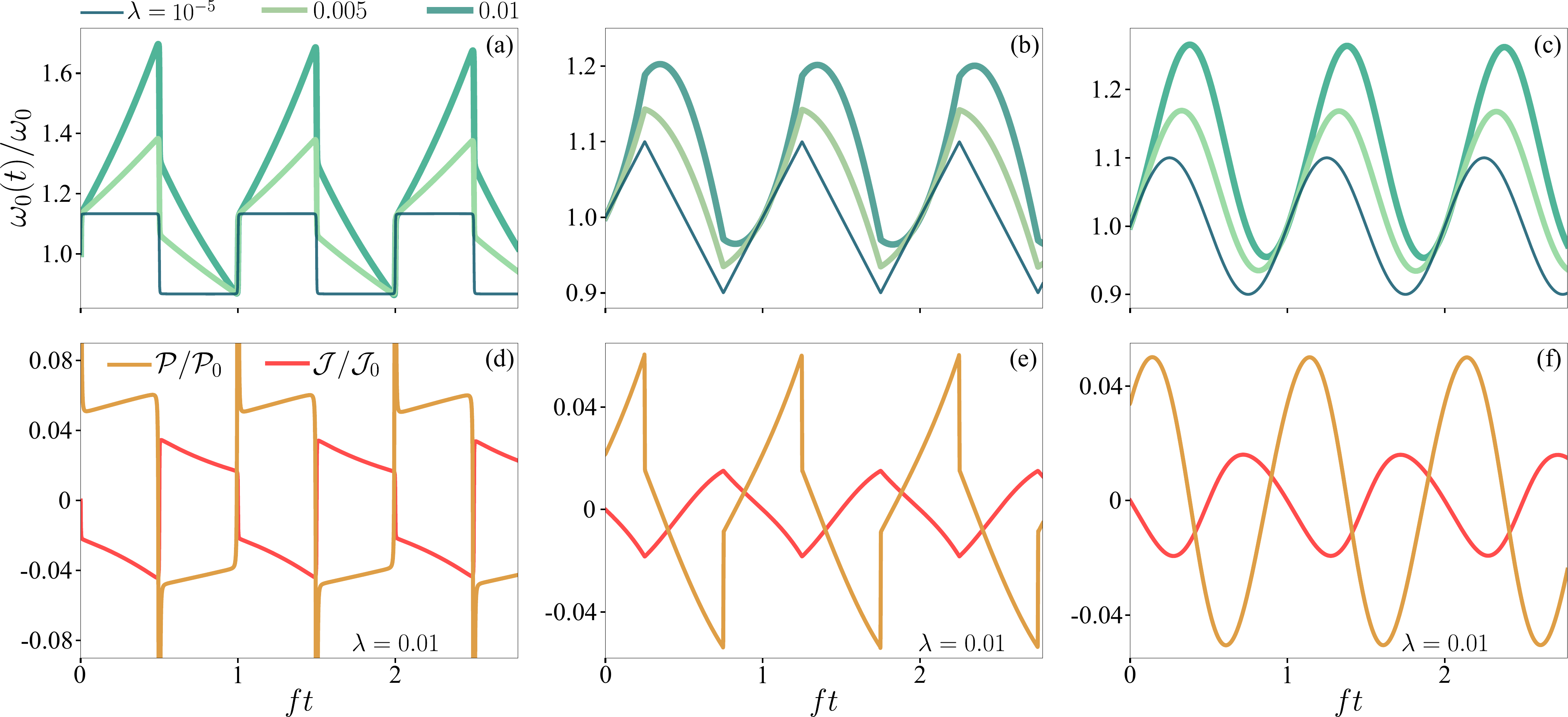}
    \caption{Driving protocol. (a) Time-dependent splitting for realizing a square-shaped temperature for different reservoir couplings and the driving frequency $f = 0.1\omega_0$, where $\omega_0=\omega_0(0)$ here is the initial splitting. For the weakest coupling, the splitting is directly given by the desired temperature, since $\omega_0(t)/\omega_0(0)=T_0(t)/T_0(0)$ according to Eq.~(\ref{eq:time-dep-temp}). (b,c) Similar results for a saw tooth and a cosine temperature. (d,e,f) Power and heat added to the qubit for $\lambda=0.01$ with $\mathcal P_0= \mathcal J_0 = f\hbar\omega_0$.}  
    \label{fig:drive} 
\end{figure*}

Figures~\ref{fig:1qubit}(d), \ref{fig:1qubit}(e), and \ref{fig:1qubit}(f) show the power and heat that are added to the qubit for the three drives and a finite reservoir coupling. We see that power is applied to (or removed from) the qubit to increase (or decrease) its temperature. This conclusion can also be reached from Eq.~(\ref{eq:q_pow}), which shows that the sign of the power is given by the change of the qubit splitting. The heat, by contrast, is given by the temperature difference between the qubit and the reservoir and vanishes in equilibrium.

\section{Driving protocol}
\label{sec:drive_protocol}

In the previous section, we saw how the qubit temperature responds to a modulation of the qubit splitting. If the coupling to the reservoir is sufficiently weak, the temperature directly follows the modulation of the splitting according to Eq.~(\ref{eq:time-dep-temp}). However, as the coupling is increased, heat starts to flow between the qubit and the reservoir, and the response of the qubit temperature becomes more complicated. One may then ask how the splitting should be modulated to realize a specific time-dependent temperature of the qubit. To this end, we consider the change in the population of the excited state  
\begin{equation}
\begin{split}
    \partial_t f(\omega_0) = \gamma(\omega_0)\Big[&n(\omega_0)(1 - f(\omega_0)) \\ - & (1+n(\omega_0)) f(\omega_0)\Big], 
\end{split}
\label{eq:driving_protocol}
\end{equation} 
which is determined by the exchange of energy between the qubit and the reservoir. This expression follows from Eq.~(\ref{eq:master_single_qubit}) by assuming that the qubit is in a thermal state. Now, given a desired time-dependent temperature, we can solve this differential equation for the time-dependent splitting that we should implement. Here, a few simple cases can easily be understood. Without the coupling to the reservoir, the right-hand side vanishes, and we find $\omega_0(t)/\omega_0(0)=T_0(t)/T_0(0)$ as in Eq.~(\ref{eq:time-dep-temp}). Moreover, if the coupling is finite, and we want the qubit to obtain the temperature of the reservoir,  $T_0(t)=T_e$, we immediately quench the qubit splitting to $\omega_0(0^+)/\omega_0(0)=T_e/T_0(0)$ and then keep the splitting fixed. Beyond these examples, we solve the differential equation numerically given a desired time-dependent temperature.

\begin{figure*}
    \centering
\includegraphics[width=0.95\textwidth]{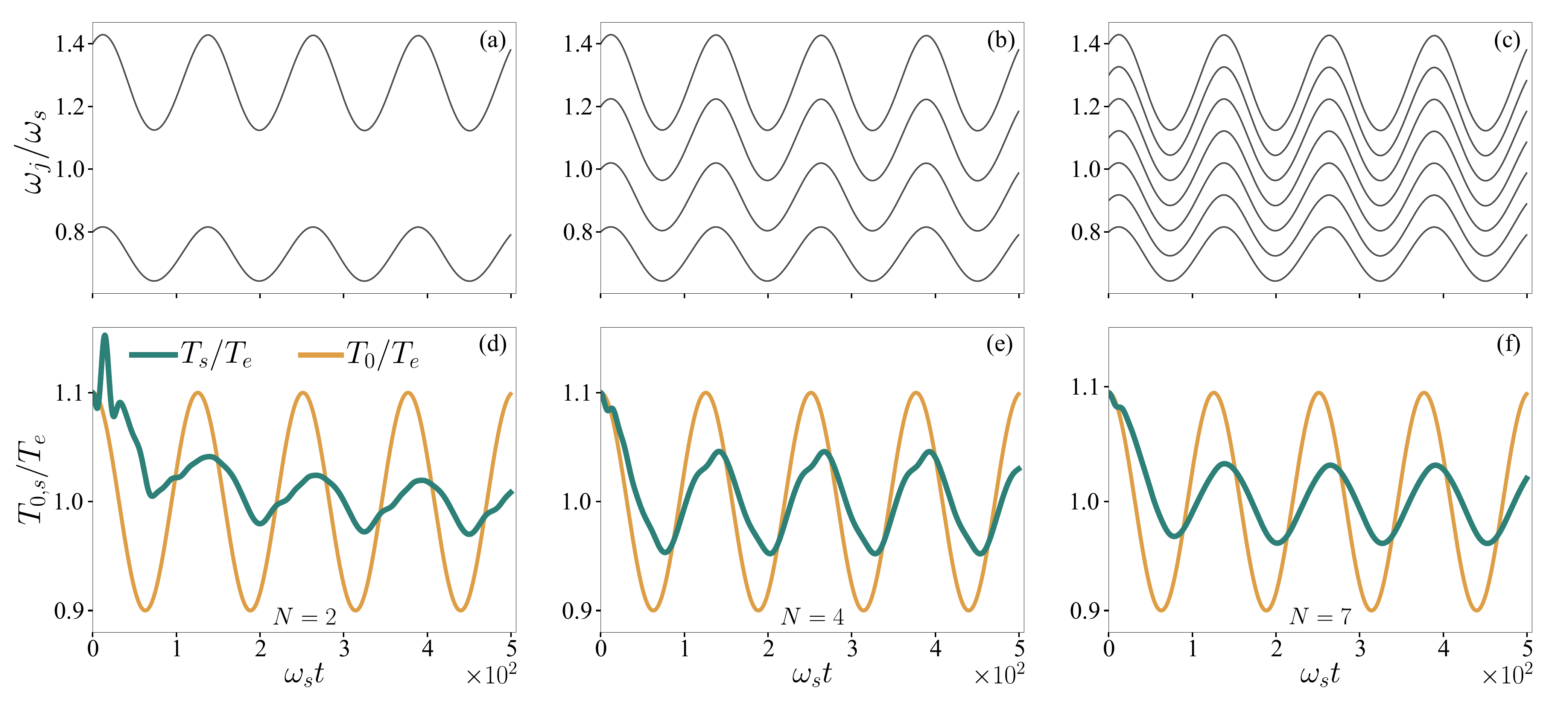}
    \caption{Cosine temperature. (a)-(c) Time-dependent splittings8c for $N=2,4,7$ ancilla qubits to realize the time-dependent temperature in the lower panels. (d)-(f) Time-dependent temperature of the ancilla qubits (yellow) and the resulting temperature of the system qubit (green). Parameters are $\lambda = 0.005$, $g = 0.15\omega_s$, and $k_B T_e = 2 \hbar \omega_s$.}
    \label{fig:cosine_effective_temperature}
\end{figure*}

Figures~\ref{fig:drive}(a), \ref{fig:drive}(b), and \ref{fig:drive}(c) show how we should modulate the splitting to realize three time-dependent temperatures of the qubit. In analogy with Fig.~\ref{fig:1qubit}, we consider a square-wave temperature, a saw-tooth temperature, and a cosine temperature for different couplings to the reservoir. For the smallest coupling, the qubit splitting again follows the time-dependent temperature according to Eq.~(\ref{eq:time-dep-temp}). By contrast, for larger  couplings, it becomes necessary to compensate for the heat that flows between the qubit and the reservoir by adding or extracting power from the qubit. The figure illustrates that it indeed is possible to change the temperature of the qubit in a controlled manner by modulating its splitting. Figures~\ref{fig:drive}(d), \ref{fig:drive}(e), and \ref{fig:drive}(f) show the power and heat that are added to the qubit for a given coupling to the reservoir. For example, in Fig.~\ref{fig:drive}(d), we see how power is added or removed to compensate for the heat flow between the qubit and the environment to keep the qubit temperature fixed. 

\section{Temperature control}
\label{sec:eff_temperatures} 

Having seen that we can control the temperature of a qubit by modulating its energy splitting, we will now explore, whether we can use one or several qubits to engineer the thermal environment of another quantum system. To this end, we consider $N$ ancilla qubits, whose energy splittings we can control, and which are coupled to a quantum system, which we take to be another qubit to keep the discussion simple. \revision{In particular, the system qubit will have a well-defined temperature during the time-dependent protocol, which would not necessarily be the case for other quantum systems. However, also for those, our scheme is useful, for example, to cool them down close to their ground state, and instead  of the temperature one could investigate how the average energy evolves with time.} 

The Hamiltonian of the coupled qubits reads
\begin{equation}
\hat H(t)=\hat H_S+\hat H_A(t)+ \hat H_{SA},
\label{eq:Hallqubits}
\end{equation}
where the qubits are described by the terms
\begin{equation}
\hat H_S=\frac{\hbar\omega_s}{2} \hat\sigma^{(s)}_z
\end{equation}
and
\begin{equation}
\hat H_A(t)=\sum_{j=1}^{N} \frac{\hbar\omega_j(t)}{2}\hat\sigma_z^{(j)},
\end{equation}
while the interaction term takes the form
\begin{equation}
    \hat H_{SA}=\hbar \sum_{j=1}^{N} g_j \left(\hat\sigma_+^{(s)} \hat\sigma_-^{(j)} + \hat\sigma_-^{(s)}\hat\sigma_+^{(j)}\right).
\end{equation}
The choice of interactions is not essential for the following discussion, and one could consider other types of interactions depending on the particular experimental setup. Moreover, for the sake of simplicity, we take the same coupling between all ancilla qubits and the system qubit and assume that it is weak, $g_j\ll\omega_s,\omega_j$. \revision{We also define $g=Ng_j$ as the sum of all couplings to the system qubit.} The ancilla qubits are again weakly coupled to a heat reservoir at the temperature~$T_e$. \revision{For now, we assume that the system qubit is only coupled to the heat reservoir via the ancilla qubits, but at the end we will consider a direct coupling between the system qubit and the heat reservoir.} The density matrix of the qubits now evolves according to the quantum master equation 
\begin{equation}
    \begin{split}
        \frac{d \hat \rho(t)}{dt} = -\frac{i}{\hbar}[\hat H(t), \hat \rho(t)] + \sum_{j = 1}^N \gamma_j(\omega_j)\mathcal{D}_j\hat \rho (t),
    \end{split} \label{eq:master}
\end{equation}
where the dissipators for each ancilla qubit take the same form as in Eq.~(\ref{eq:dissipator}), and the reservoir coupling is assumed to be the same for all of them. The derivation of the quantum master equation is presented in Appendix~\ref{appA}, while a numerical scheme for solving it is described in Appendix~\ref{appC}.

\begin{figure*}
	\centering
	\includegraphics[width=0.95\textwidth]{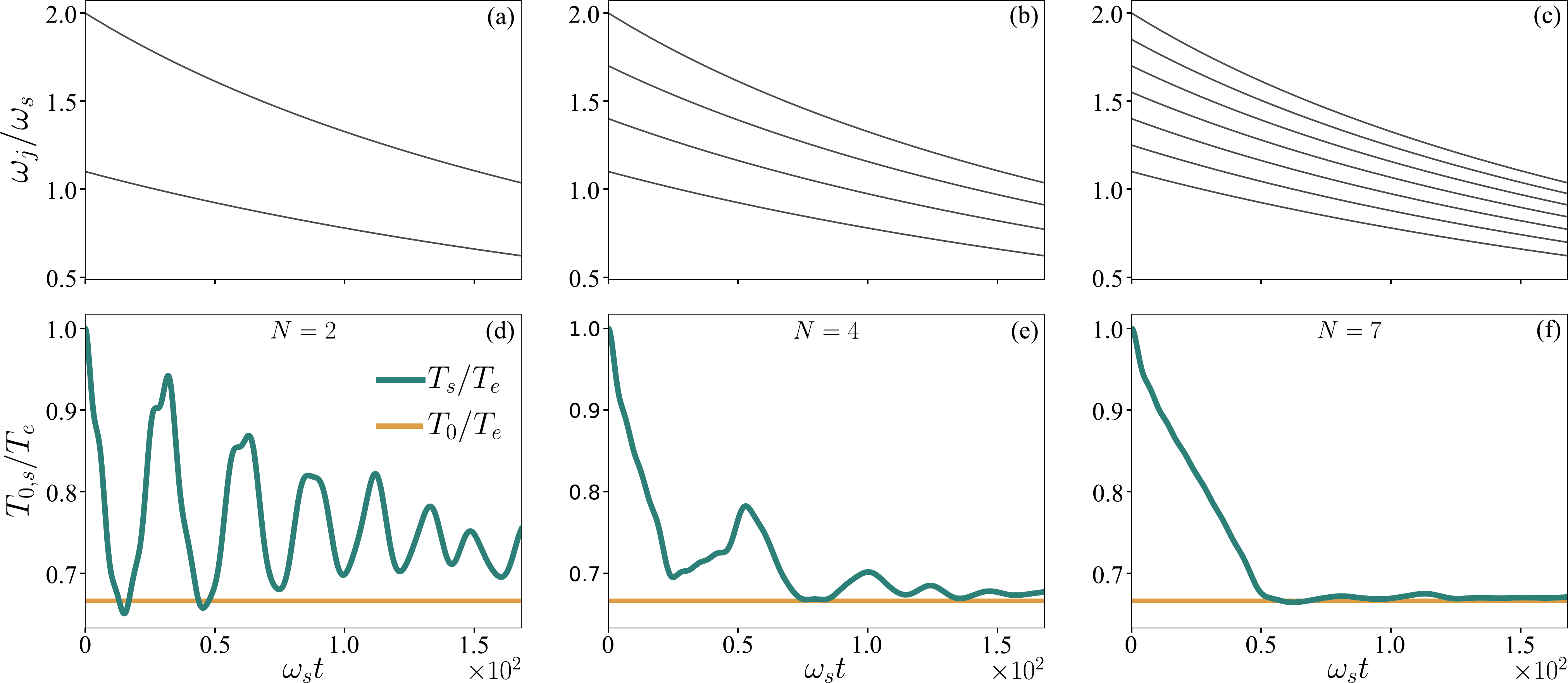}
	\caption{Cooling. (a)-(c) Time-dependent splittings for $N = 2, 4, 7$ ancilla qubits to reach the constant temperature $k_B T_0 = \hbar \omega_s$ below the temperature of the reservoir, $k_B T_e = 1.5 \hbar \omega_s$. (d)-(f) Time-dependent temperature of the system qubit (green), and the constant temperature of the ancilla qubits (orange). The other parameters  are $\lambda = 0.001$ and $g = 0.2\omega_s$.}
	\label{fig:constant_effective_temperature}
\end{figure*}

We can now implement a time-dependent temperature for the ancilla qubits by solving Eq.~(\ref{eq:driving_protocol}) for their splittings without the coupling to the system qubit. We then consider the time-evolution of the coupled system and ancilla qubits using the time-dependent splittings. With a weak coupling between the system and ancilla qubits, the ancilla qubits are to a good approximation described by the thermal state in Eq.~(\ref{eq:equi_den}) with the time-dependent temperature inserted. It is therefore conceivable that the system qubit will evolve as if the ancilla qubits were in thermal equilibrium with a reservoir that had the time-dependent temperature and not the actual one.

Figure~\ref{fig:cosine_effective_temperature} shows the temperature of the system qubit as it is exposed to the time-dependent temperature of the ancilla qubits. We  first consider a cosine temperature modulation of the ancilla qubits. The first row shows the time-dependent splittings for $N=2,4,$ and $7$ qubits, while the corresponding temperatures of the ancilla qubits are shown in the second row together with the resulting temperature of the system qubit. To begin with, the temperature of all qubits is higher than the reservoir temperature. The initial splittings of the ancilla qubits are chosen by having in mind that energy transfer is a resonant process, which is most efficient around $\omega_j(t) \simeq \omega_s$.  The temperature of the system qubit clearly responds to the time-dependent temperature of only two ancilla qubits; however, the time-evolution is somewhat irregular. By contrast, as the number of ancilla qubits is increased, the behavior becomes smoother, and already with seven ancilla qubits, the temperature of the system qubit quickly adapts to that of the ancilla qubits. Indeed, after a short transient, the temperature of the system qubit synchronizes with the temperature of the ancilla qubits, however, with a reduction of the amplitude and a phase-lag due to the finite time for the system qubit to equilibrate with the ancilla qubits. 

Next, we explore whether it is possible to cool the system qubit using the ancilla qubits. Initially, all qubits have the temperature of the reservoir. We then immediately reduce the splittings of the ancilla qubits as $\omega_0(0^+)/\omega_0(0)=T_0(0^+)/T_e$, which reduces their temperature to $T_0(0^+)$, which is well below the temperature of the reservoir. Heat now starts to flow from the reservoir into the ancilla qubits. However, to keep their temperature constant, we further  reduce their energy splittings as shown in Figs.~\ref{fig:constant_effective_temperature}(a), \ref{fig:constant_effective_temperature}(b), and \ref{fig:constant_effective_temperature}(c) with results for $N=2,4,$ and 7 ancilla qubits. In Fig.~\ref{fig:constant_effective_temperature}(d), a cooling effect is already observed with only two qubits; however, the temperature of the system qubit oscillates, and the cooling process is far from being ideal. By contrast, with an increasing number of ancilla qubits, as in Fig.~\ref{fig:constant_effective_temperature}(e), the cooling process becomes smoother, and with seven ancilla qubits, the temperature of the system qubit quickly reaches that of the ancilla qubits as we see in Fig.~\ref{fig:constant_effective_temperature}(f). To maintain the constant temperature of the ancilla qubits, it is necessary to compensate for the heat that flows into them from the reservoir. Therefore, the qubit splittings must continuously be reduced; however, eventually, the ancilla qubits and the system qubit will no longer be in resonance, and the cooling process becomes inefficient. At that point, the ancilla qubits can be reset by increasing their splittings, and the cooling process can repeat. 

\begin{figure}
    \centering
\includegraphics[width=0.9\columnwidth]{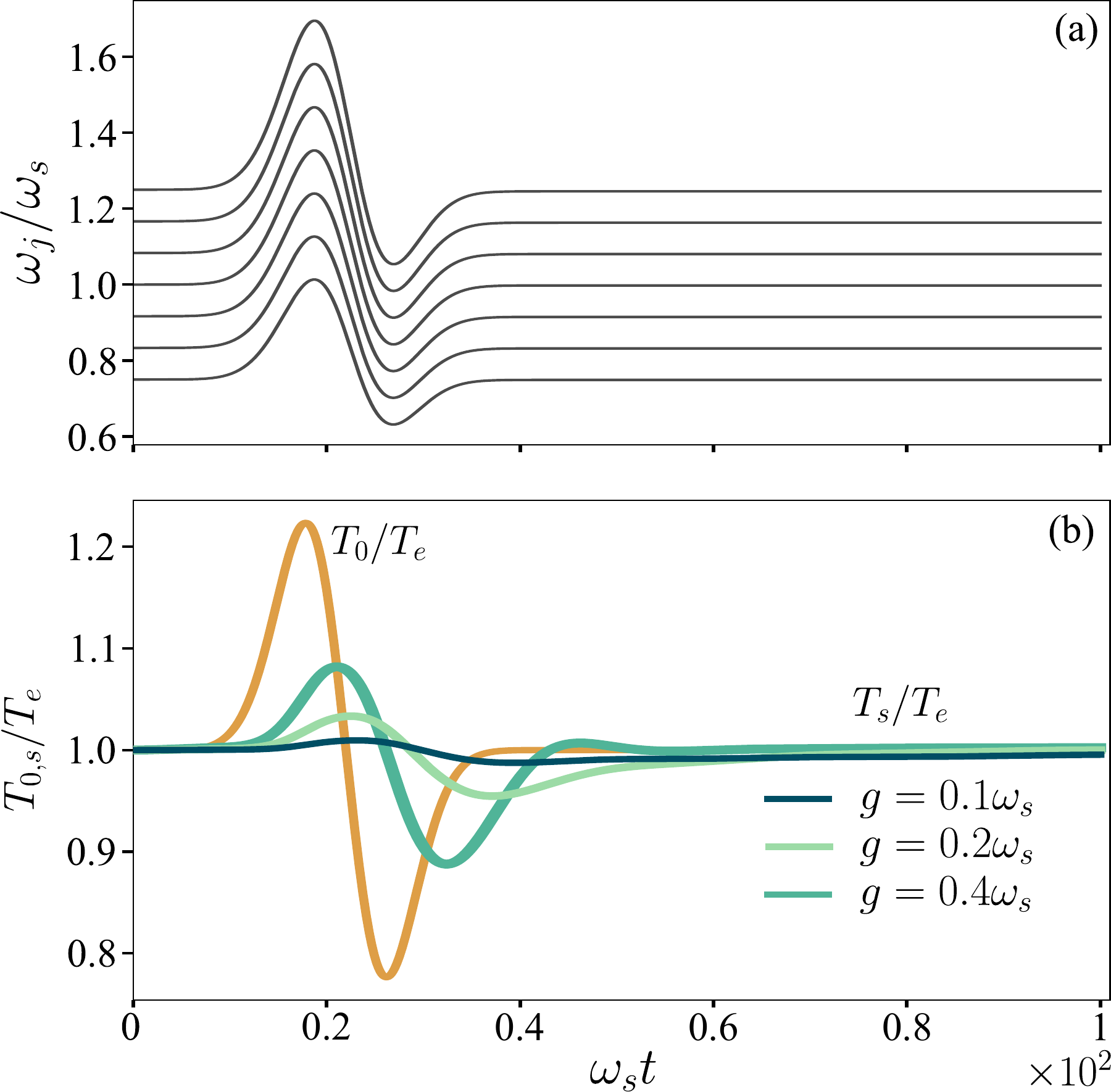}
    \caption{Temperature pulse. (a) Time-dependent splittings for $N = 7$ ancilla qubits to realize the temperature pulse $T_0(t)=T_e+\Delta T_0 (e^{-(t-t_0)^2/2\sigma^2}-e^{-(t-t_0-\Delta t_0)^2/2\sigma^2})$. (d) Temperature of the ancilla qubits and the system qubit for different couplings. The other parameters are $k_B T_e = 1.5 \hbar \omega_s$, $k_B \Delta T_0=0.6\hbar\omega_s$, $\omega_s t_0 = 20$,  $\omega_s \Delta t_0 = \omega_s \sigma = 4$ and $\lambda = 0.01$.} 
    \label{fig:discrete_pulse}
\end{figure}

As our last application, we consider a thermal excitation of the system qubit by a temperature pulse. Such short excitations may be useful to probe the dynamic thermal response of a quantum system in the time-domain and to determine its thermal spectral properties. In Fig.~\ref{fig:discrete_pulse}(a), we show how the energy splittings of the ancilla qubits are modulated in time to generate the temperature pulse in Fig.~\ref{fig:discrete_pulse}(b) with a fixed number of qubits. In Fig.~\ref{fig:discrete_pulse}(b), we also show the time-dependent temperature of the system qubit for three different couplings to the ancilla qubits, as the system qubit responds to the temperature pulse. Here, we observe a finite response time of the system qubit, which becomes shorter as the coupling is increased. Also, with an increasing coupling, the flow of heat is enhanced, and the amplitude of the thermal response gets larger. 

\section{Experimental perspectives}
\label{sec:exp_persp}

Finally, we discuss the perspectives of realizing our proposal experimentally. Our setup is based on several qubits, whose energy splittings can be modulated in time to control the thermal environment of another quantum system. Possible implementations could be based on superconducting flux qubits~\cite{Krantz2019,Anferov2024}, charge qubits~\cite{Makhlin:2001}, or spin qubits~\cite{Hanson2007,Burkard:2023}, but the use of other two-level systems can also be envisioned. To provide a specific example, we here discuss a potential implementation based on spin qubits, taking inspiration from the experiments of Refs.~\cite{Ono:2019,Ono:2020}, where the energy splitting of a single spin qubit was modulated in time using a square wave, a saw-tooth, and a cosine drive. Typical energy splittings of spin qubits are in the range of one to hundreds of millielectron-volts, depending in part on the host material, the confining potential, and the applied magnetic field. Such splittings correspond to frequencies in the gigahertz-regime and temperatures in the subkelvin-range. Taking as an example $\hbar\omega_s=10$~$\mu$eV (or $\omega_s=2.4$~GHz), we find that the quantum system in Fig.~\ref{fig:constant_effective_temperature} would be cooled from $T_e=1.5\hbar\omega_s/k_B=180$~mK to $T_0=\hbar\omega_s/k_B=120$~mK in about 20~ns. Such temperatures and time scales are accessible with current technology, and our estimates would also apply to other types of solid-state qubits. 

\begin{figure}
    \centering
    \includegraphics[width=0.5\columnwidth]{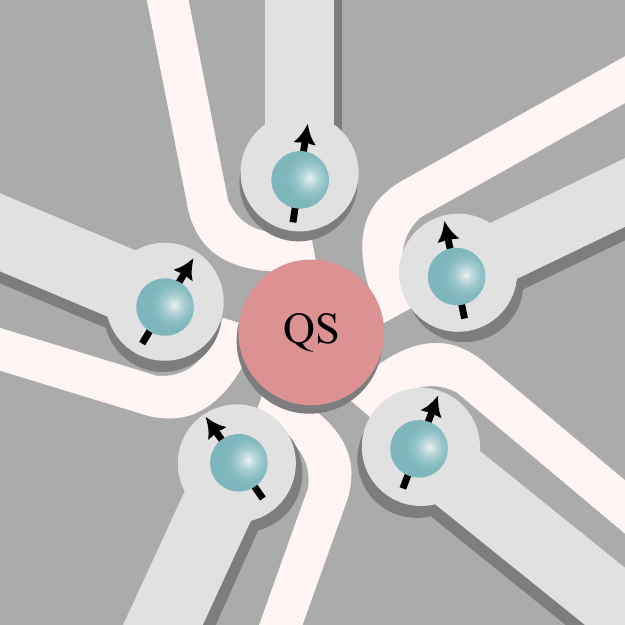}
    \caption{Potential experimental setup. Five quantum dots each host an electron or a hole spin qubit, whose energy splittings can be controlled by a magnetic field or by using electric gate pulses combined with a strong spin-orbit coupling as in  Ref.~\cite{Hendrickx:2021}. Each qubits is coupled to the central quantum system (QS), whose thermal environment we wish to control.} 
    \label{fig:exp-setup}
\end{figure}

\begin{figure}
	\centering
	\includegraphics[width=0.9\columnwidth]{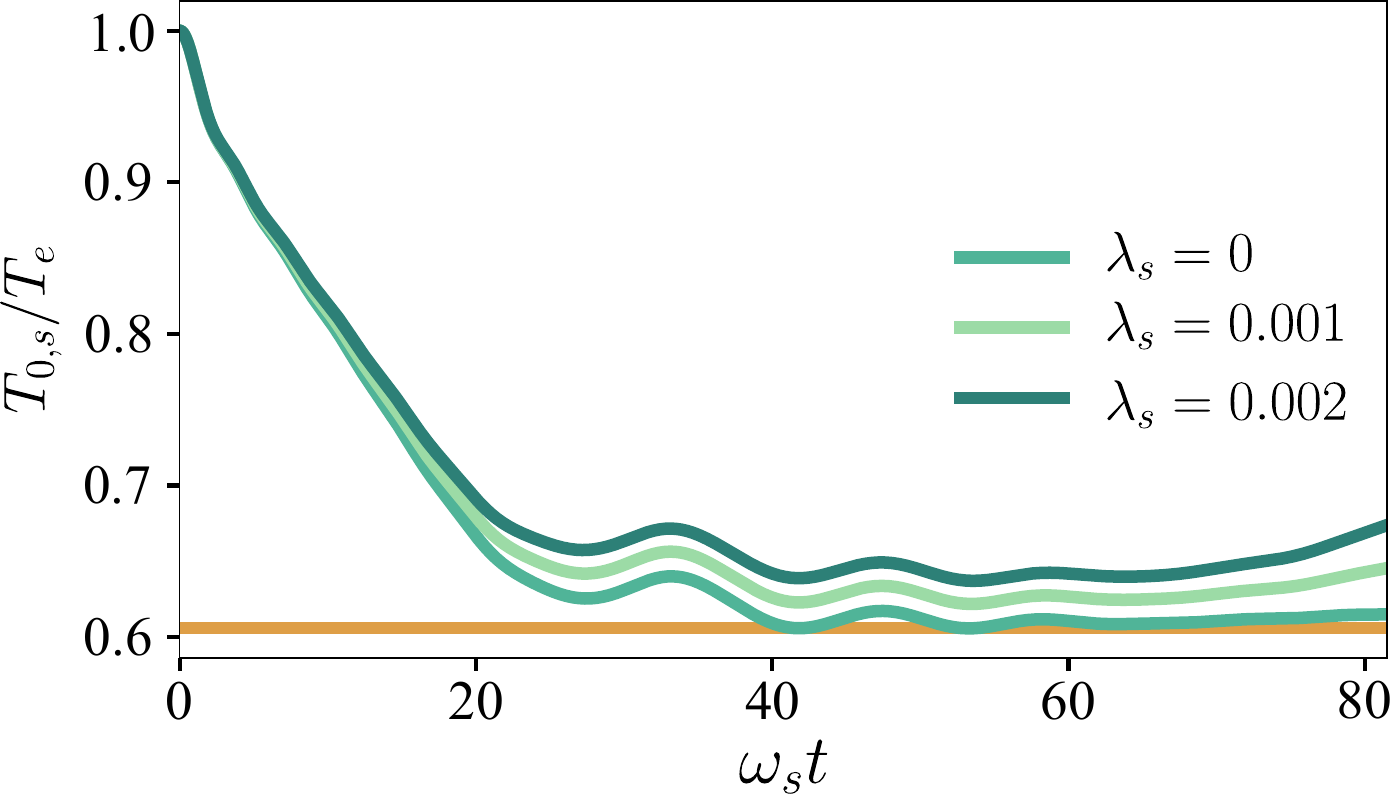}
	\caption{\revision{Direct coupling between system qubit and reservoir. We show the time-dependent temperature of $N = 7$ ancilla qubits (orange) and  the system qubit (green) for three different couplings between the system qubit and the heat reservoir, denoted by $\lambda_s$. The other parameters  are $\lambda = 0.001$, $g = 0.2\omega_s$, and $k_B T_e = 1.65 \hbar \omega_s$. The results for $\lambda_s=0$ correspond to Fig.~\ref{fig:system}(c) with no coupling to the environment.}} 
	\label{fig:env-coup}
\end{figure}

In practice, one may consider a setup as outlined in Fig.~\ref{fig:exp-setup}, where several spin qubits are positioned around a quantum system at the center, whose local thermal environment we wish to control. In our calculations, we have taken the same coupling strength between the qubits and the quantum system. However,  our proposal should work equally well with different couplings, since it is mainly based on the heat flow that is generated by temperature differences, and such thermodynamic processes should not be very sensitive to the exact values of the couplings. Also, the quantum system may itself exchange heat with the host environment, which would only have a minor influence as long as heat is predominately flowing between the quantum system and the qubits. \revision{Indeed,  Fig.~\ref{fig:env-coup} shows how the system qubit can be cooled despite a direct coupling to the heat reservoir, even if is twice as large as the coupling of the ancilla qubits to the reservoir.} Finally, the ability to modulate qubit splittings accurately and fast directly translates into an accurate and fast control of the local temperature. Thus, our estimates indicate that qubits can be used for fast, accurate, and local control of the temperature in solid-state systems. 

\section{Conclusions and outlook}
\label{sec:conclusion}

We have presented a proposal for achieving fast, accurate, and local temperature control using qubits. To this end, work is performed on the qubits by modulating their energy splittings, and a desired time-dependent temperature can be realized by compensating for the heat that flows between the qubits and their environment. \revision{Using less than 10 qubits}, it is possible to control the thermal environment of another quantum system, which can be heated or cooled by the qubits. Our proposal is based on basic thermodynamic principles, and we expect that it is not very dependent on the particular details of our model or the specific choice of parameters. Also, the proposal can potentially be realized using different types of qubits, including superconducting flux qubits, charge qubits, or spin qubits, which can now \revision{be} fabricated and manipulated with exquisite control. As an example, we have shown how a quantum system at subkelvin temperatures can be cooled on a nanosecond timescale. 

Our work can be extended in many directions. As the next step, it would be important to adapt our ideas to specific physical setups and qubit architectures. In addition to practical realizations of our proposal for temperature control,  several other applications can be envisioned. For instance, in the context of dynamic heat transport, one may explore thermotronic effects by \revision{by using several qubits to} inject power on one side of a quantum system and investigate the output of heat on the other side~\cite{Portugal:2021}. It may also be possible to use qubits to locally cool a quantum circuit and thereby create a heat sink, which would allow for unwanted waste heat to be removed. Finally, one could realize geometric pumping of heat \revision{using qubits} and observe the associated Berry phase~\cite{Ren:2010}. These, and other ideas, we leave for future work.

\begin{acknowledgments}
We thank F.~Brange, J.~Chen, K.~Funo, N.~Lambert, P.~Menczel, and F.~Nori for useful discussions and acknowledge the support from the Nokia Industrial Doctoral School in Quantum Technology, the Research Council of Finland through the Finnish Centre of Excellence in Quantum Technology
(Grant No.~352925), and the Japan Society for the Promotion of Science through an Invitational Fellowship for Research in Japan. 
\end{acknowledgments}

\appendix
\section{Quantum master equation}
\label{appA}	
Here, we derive the quantum master equation in Eq.~\eqref{eq:master} following standard procedures such as those described in Refs.~\cite{breuer2002theory,Albash:2012,Rivas:2012,Portugal2022}. The derivation is similar to the one for a time-independent Hamiltonian, and for the sake of completeness we include it here to show how it can be adapted to our specific time-dependent situation. 

The Hamiltonian for our setup takes the form
\begin{equation}
		\hat H_\mathrm{tot}(t)=\hat H(t)+\hat H_B+\hat H_I,
\end{equation}
where $\hat H(t)$ is the time-dependent Hamiltonian in Eq.~(\ref{eq:Hallqubits}) for the coupled qubits. In addition, the ancilla qubits are coupled to an external reservoir, which is described as a collection of harmonic oscillators, 
 \begin{equation}
     \hat H_B = \sum_l \hbar \omega_l \hat a_l^\dagger \hat a_l,
 \end{equation}
while the interaction Hamiltonian reads
\begin{equation}
    \hat H_I = \sum_{j = 1}^N \frac{\hbar}{2} \hat \sigma_x^{(j)}\otimes  \sum_l(\kappa_l \hat a_l + \kappa_l^* \hat a_l^\dagger).
\end{equation}

The density matrix of the full setup evolves according to the Liouville-von Neumann equation, 
\begin{equation}
    \frac{d \hat \rho_\mathrm{tot}(t)}{dt} = -\frac{i}{\hbar}[\hat H_\mathrm{tot}(t), \hat \rho_\mathrm{tot}(t)]. 
\end{equation}
Next, we switch to the interaction picture with respect to $\hat H(t) + \hat H_B$ by defining the unitary operators 
\begin{equation}
\begin{split}
\hat U_0(t,t_0)&=\hat{T}\left\{ e^{-i\int_{t_0}^tds[\hat H(s)+\hat H_B]/\hbar}\right\}\\
&=\hat{T}\left\{ e^{-i\int_{t_0}^tds \hat H(s)/\hbar}\right\}\otimes e^{-i\hat H_B(t-t_0) /\hbar}\\
&\equiv \hat U(t,t_0)\otimes \hat U_B(t,t_0),
\end{split}
\end{equation}
where $\hat{T}$ is the time-ordering operator. In the interaction picture, the density matrix reads 
\begin{equation}
\tilde \rho_\mathrm{tot}(t)= \hat U_0^\dagger(t,0) \hat \rho_\mathrm{tot}(t) \hat U_0(t,0),
\end{equation} 
and the Hamiltonian becomes 
\begin{equation}
\tilde H_I(t)=\hat U^\dagger_0(t,0) \hat H_I\hat U_0(t,0).
\end{equation}
We then find 
\begin{equation}
        \frac{d \tilde \rho_\mathrm{tot}(t)}{dt} = -\frac{i}{\hbar}[\tilde H_I(t), \tilde \rho_\mathrm{tot}(t)]. 
        \label{eq:eom-rhoT}
\end{equation}
which can be integrated and iterated once to yield 
\begin{equation}
\begin{split}
\tilde \rho_\mathrm{tot} (t) = &\tilde \rho_\mathrm{tot}(0) - \frac{i}{\hbar} \int_0^t ds [\tilde H_I(s), \tilde 
\rho_\mathrm{tot}(s)]\\
=&\tilde \rho_\mathrm{tot}(0) - \frac{i}{\hbar} \int_0^t ds [\tilde H_I(s), \tilde 
\rho_\mathrm{tot}(0)]\\
&-\frac{1}{\hbar^2}\int_0^t ds \int_0^s ds' [\tilde H_I(s), [\tilde H_I(s'), \tilde \rho_\mathrm{tot}(s')]].
\end{split}
\end{equation}
Next, we trace out the reservoir degrees of freedom and consider the reduced density matrix of the qubits
\begin{equation}
\begin{split}
    \tilde \rho(t)&=\mathrm{tr}_B\{\hat U_0^\dagger(t,0) \hat \rho_\mathrm{tot}(t) \hat U_0(t,0)\}\\
    &=\hat U^\dagger(t,0)\mathrm{tr}_B\{\hat U_B^\dagger(t,0) \hat \rho_\mathrm{tot}(t) \hat U_B(t,0)\}\hat U(t,0)\\
    &=\hat U^\dagger(t,0)\hat \rho(t) \hat U(t,0),
\end{split}
\end{equation}
where the cyclic property of the trace was used in the last step. Combining these expressions, we find
\begin{equation}
\begin{split}
    \frac{d \tilde \rho(t)}{dt} =& -\frac{i}{\hbar}\text{tr}_B\{[\tilde H_I(t), \tilde \rho_\mathrm{tot}(0)]\} \\
    &- \frac{1}{\hbar^2}\int_0^t ds \text{ tr}_B\{[\tilde H_I(t), [\tilde H_I(s), \tilde \rho_\mathrm{tot}(s)]]\}.
\end{split}
\label{eq:2ndOrdereom_trace}
\end{equation}

To proceed, we now make the Born approximation by assuming that there are no correlations between the environment and the qubits. Specifically, we assume that 
\begin{equation}
    \tilde \rho_\mathrm{tot}(t)\simeq\tilde \rho(t)\otimes \hat \rho_B,
\end{equation}
where $\hat\rho_B$ is the thermal density matrix of the reservoir at the temperature $T_e$. The first term in Eq.~\eqref{eq:2ndOrdereom_trace} vanishes because of this assumption, 
\begin{equation}
    \mathrm{tr}_B\{[\tilde H_I(t),\tilde \rho_\mathrm{tot}(0)]\} =\mathrm{tr}_B\{[\tilde H_I(t),\tilde \rho(0)\otimes \hat \rho_B]\} = 0.
\end{equation}
In addition, we  make the Markov approximation by replacing $\tilde\rho(s)$ by $\tilde\rho(t)$ in Eq.~(\ref{eq:2ndOrdereom_trace}), such that the time-evolution of the density matrix becomes local in time. We then make a change of integration variable, $s\rightarrow t - s$, and extend the  upper integration limit  to infinity. We then obtain the integro-differential equation
\begin{equation}
    \frac{d \tilde \rho(t)}{dt} = - \frac{1}{\hbar^2}\int_0^\infty ds \text{ tr}_B[\tilde H_I(t), [\tilde H_I(t - s), \tilde \rho(t) \otimes \hat\rho_B]]. 
    \label{eq:master_eq_formal}
\end{equation}

Next, we transform the interaction Hamiltonian to the interaction picture, starting with the Pauli matrices. Here, we take the coupling between the system qubit and the ancilla qubits to be weak, $g_j \ll \omega_s, \omega_j$, such that the ladder operators $\hat \sigma_{+}^{(j)}$ and $ \hat \sigma_{-}^{(j)}$ approximately fulfill that 
\begin{equation}
[\hat H(t),\hat \sigma_{-}^{(j)}] = - \hbar \omega_j(t) \hat \sigma_{-}^{(j)}
\end{equation}
and 
\begin{equation}
[\hat H(t),\hat \sigma_{+}^{(j)}] = - \hbar\omega_j(t) \hat \sigma_{+}^{(j)}.
\end{equation}
We then find 
\begin{equation}
        \tilde \sigma_{+}^{(j)}(t)= \hat U^\dagger (t, 0) \hat \sigma_{+}^{(j)} \hat U(t, 0)\simeq e^{i \int_0^t \omega_j(s') ds'} \hat \sigma_{+}^{(j)},
\end{equation}
and
\begin{equation}
        \hat \sigma_{-}^{(j)}(t) = \hat U^\dagger (t, 0) \hat \sigma_{+}^{(j)} \hat U(t, 0) \simeq e^{-i \int_0^t \omega_j(s') ds'} \hat \sigma_{+}^{(j)},
\end{equation}
We also assume that $\hat H(t)$ does not change considerably on the timescale over which reservoir correlations decay,~$\tau_B$. For  $s\lesssim\tau_B$, we can then approximate 
\begin{equation}
\hat U(t-s, 0) = \hat U^\dagger(t, t-s) \hat U(t, 0) 
\simeq   e^{i \hat H(t) s/\hbar}\hat U(t, 0).
\end{equation}
Therefore, we have
\begin{equation}
    \begin{split}
        \tilde \sigma_{+}^{(j)}(t - s) &= \hat U^\dagger (t - s, 0) \hat  \sigma_{+}^{(j)} \hat U(t - s, 0) \\
        &= e^{i \int_0^t \omega_j(s') ds'} e^{-i \omega_j(t) s} \hat \sigma_{+}^{(j)}, 
\end{split}
\end{equation}
and
\begin{equation}
    \begin{split}
        \tilde \sigma_{-}^{(j)}(t - s) &= \hat U^\dagger (t - s, 0) \hat \sigma_{-}^{(j)} \hat U(t - s, 0)\\
        &= e^{-i \int_0^t \omega_j(s') ds'} e^{i \omega_j(t) s} \hat \sigma_{-}^{(j)}.
    \end{split}
\end{equation}

Next, we insert these expressions into Eq.~\eqref{eq:master_eq_formal} and make the secular wave approximation. To this end, we note that we have $\omega_{j\neq k}(t)\neq \omega_k(t)$ at all times. We then remove all fast-rotating terms to obtain
\begin{equation}
    \begin{split}
        \frac{d \tilde \rho(t)}{dt} = \frac{1}{4} \sum_{j = 1}^N \Big(&\Gamma(\omega_j(t))(\hat \sigma_{-}^{(j)} \tilde \rho(t) \hat  \sigma_{+}^{(j)} - \hat  \sigma_{+}^{(j)} \hat \sigma_{-}^{(j)} \tilde \rho (t)) \\
        +& \Gamma(-\omega_j(t))(\hat  \sigma_{+}^{(j)} \tilde \rho(t) \hat \sigma_j^- - \hat \sigma_{-}^{(j)} \hat \sigma_j^+ \tilde \rho (t)) \\+& 
        \Gamma^*(\omega_j(t))(\hat \sigma_{-}^{(j)} \tilde \rho(t) \hat  \sigma_{+}^{(j)} - \tilde \rho (t) \hat  \sigma_{+}^{(j)} \hat \sigma_{-}^{(j)} )\\
         +& \Gamma^*(-\omega_j(t))(\hat \sigma_{+}^{(j)} \tilde \rho(t) \hat \sigma_{-}^{(j)} - \tilde \rho (t)\hat \sigma_{-}^{(j)} \hat \sigma_{+}^{(j)}) \Big),
    \end{split}
\end{equation}
where we have defined
\begin{equation}
\Gamma(\omega) = \int_0^\infty ds e^{i \omega s} \text{tr}_B \{ \tilde B(t) \tilde B(t-s) \hat \rho_B \}
\end{equation}
in terms of the operator of the bath,
\begin{equation}
\tilde B(t) = \sum_l(\kappa_l \hat a_l e^{-i \omega_l t} + \kappa_l^* \hat a_l^\dagger e^{i \omega_l t}) .
\end{equation}
The correlation function above only depends on the time difference, such that we can define $C(s)=C(t,t-s)$ with
\begin{equation}
C(s)= \text{tr}_B \{ \tilde B(t) \tilde B(t-s) \hat \rho_B \},
\end{equation}
We also decompose the correlation function into real and imaginary parts as 
\begin{equation}
\Gamma(\omega) = \frac{1}{2}\mu(\omega) + i\eta(\omega).
\end{equation}
Using this decomposition and transforming operators back into the Schr\" odinger picture, we find
\begin{equation}
    \begin{split}
        \frac{d \hat \rho(t)}{dt} = &-\frac{i}{\hbar}[\hat H(t) + \hat H_{LS}(t), \hat \rho(t)] \\
        +  &\sum_{j = 1}^N \left(\gamma_{-}(\omega_j(t))(\hat \sigma_{-}^{(j)} \hat \rho(t) \hat \sigma_{+}^{(j)} - \{\hat \sigma_{+}^{(j)} \hat \sigma_{-}^{(j)}, \hat \rho (t)\}/2)\right. \\
        +& \left.
         \gamma_{+}(\omega_j(t))(\hat \sigma_{+}^{(j)} \hat \rho(t) \hat \sigma_{-}^{(j)} - \{\hat \sigma_{-}^{(j)} \hat \sigma_{+}^{(j)}, \hat \rho (t)\}/2)\right),
    \end{split}
\end{equation}
where we have introduced the rates
\begin{equation}
\gamma_{\pm}(\omega) = \frac{1}{4}\mu(\mp \omega),
\end{equation}
and the Hamiltonian for the Lamb shift reads 
\begin{equation}
\hat H_{LS} = \sum_{j = 1}^N\left(\eta(\omega_j(t))\hat\sigma_{+}^{(j)}\hat\sigma_{-}^{(j)} + \eta(-\omega_j(t))\sigma_{-}^{(j)}\sigma_{+}^{(j)}\right).
\end{equation}
The Hamiltonian for the Lamb shift commutes with the qubit Hamiltonian $\hat H(t)$ for $g_j \ll \omega_s, \omega_j$  and only leads to a renormalization of the energy levels. The Lamb shift is of second order in the reservoir coupling and is therefore small, such that we can ignore it in the following. 

Next, we evaluate the reservoir correlation function
\begin{equation}
\begin{split}
C(s) = &\sum_{l,k}  \kappa_l \kappa_k e^{-i \omega_l s} \langle \hat a_l \hat  a_k \rangle  + \kappa_l \kappa_k^* e^{- i \omega_l s}\langle \hat  a_l \hat 
 a_k^{\dagger} \rangle\\
&+ \kappa_l^* \kappa_k e^{i \omega_l s} \langle \hat  a_l^{\dagger} \hat  a_k \rangle + \kappa_l^* \kappa_k^* e^{i \omega_l s} \langle \hat  a_l^{\dagger} \hat 
 a_k^{\dagger}\rangle.
\end{split}
\end{equation}
We then use that
\begin{equation}
\begin{split}
\langle \hat a_l \hat a_k \rangle &= 0,\\
\langle \hat a_l^{\dagger} \hat a_k^{\dagger} \rangle & = 0,\\
\langle \hat a_l^{\dagger} \hat a_k \rangle &=  n(\omega_l)\delta_{lk},\\
\langle \hat a_l \hat a_k^{\dagger} \rangle &=  (1+n(\omega_l))\delta_{lk}
\end{split}
\end{equation}
where $n(\omega)$ is the Bose-Einstein distribution with the temperature of the reservoir. We then find 
\begin{equation}
C(s) = \sum_l |\kappa_l|^2 \left(e^{-i \omega_l s} (1+n(\omega_l)) + e^{i \omega_l s} n(\omega_l)\right).
\end{equation}
We write the spectral density of the reservoir as
\begin{equation}
J(\omega) = \sum_l |\kappa_l|^2 \delta(\omega - \omega_l),
\end{equation}
such that 
\begin{equation}
C(s) = \int_0^{\infty} d\omega J(\omega) \left(e^{-i \omega s}(1+n(\omega)) + e^{i \omega s}n(\omega)\right).
\end{equation}
To proceed, we take an ohmic spectral density
\begin{equation}
J(\omega) = \lambda \omega e^{-\omega/\omega_c},
\end{equation}
where $\lambda$ is the coupling strength, and $\omega_c$ is a high-frequency cut-off. We then find
\begin{equation}
    \gamma_{-}(\omega) = \frac{1}{4}\int_{-\infty}^{+\infty} ds e^{i\omega s} C(s)= \frac{\pi}{2}  \lambda \omega e^{-\omega/\omega_c} (1+n(\omega))
\end{equation}
as well as
\begin{equation}
    \gamma_+(\omega) =  \frac{\pi}{2}\lambda \omega e^{-\omega/\omega_c}n(\omega).
\end{equation}
By combining all of these expressions, we finally arrive at the quantum master equation in Eq.~\eqref{eq:master}. The quantum master equation for a single qubit in Eq.~(\ref{eq:master_single_qubit}) follows as a special case by setting the coupling between the system qubit and the ancilla qubits to zero, $g=0$.

\section{Numerical time-evolution}
\label{appC} 
The quantum master equation~\eqref{eq:master} is of the form
\begin{equation}
\frac{d }{dt} \hat{\rho}(t)= \mathcal{L}(t) \hat{\rho}(t),
\end{equation}
where the time-dependent Liouvillian $\mathcal{L}(t)$ generates the time-evolution. Formally, its solution reads
\begin{equation}
\hat{\rho}(t) = \hat{T}\left\{ e^{\int_{t_0}^t ds \mathcal{L}(s)}\right\}\hat{\rho}(0),
\end{equation}
where $\hat{\rho}(0)$ is the initial state of the qubits. Numerically, we can take small time steps of length $\Delta t$ and propagate the density matrix forward in time as
\begin{equation}
\hat{\rho}(t+\Delta t) \simeq e^{ \mathcal{L}(t)\Delta t}\hat{\rho}(t).
\label{eq:time_prop}
\end{equation}
By contrast, simple Euler steps of the form
\begin{equation}
\hat{\rho}(t+\Delta t) \simeq \left(1+\mathcal{L}(t)\Delta t\right)\hat{\rho}(t)
\end{equation}
become numerically unstable as we have found. 

For the numerical implementation, one may reshape the density matrix into a vector, while the Liouvillian becomes a matrix~\cite{Landi:2024}. This approach works well for a few qubits. However, with $N=5$ qubits, the Liouvillian already contains $2^{4N}\simeq10^6$ matrix elements, while for $N=10$, the number exceeds $10^{12}$, although many are zero. Still, even with sparse matrices, memory constraints and the computational time become an issue.

To circumvent this problem, we instead employ a Crank-Nicolson scheme by rewriting Eq.~(\ref{eq:time_prop}) as
\begin{equation}
e^{ -\mathcal{L}(t)\Delta t/2}\hat{\rho}(t+\Delta t) \simeq e^{ \mathcal{L}(t)\Delta t/2}\hat{\rho}(t)
\end{equation}
and expanding both exponential functions to first order in the time step. We then invert the term that appears on the left hand-side and find
\begin{equation}
\hat{\rho}(t+\Delta t) \simeq (1-\mathcal{L}(t)\Delta t/2)^{-1}(1+\mathcal{L}(t)\Delta t/2)\hat{\rho}(t).
\end{equation}
Next, we make use of the Neumann series,
\begin{equation}
(1 - T)^{-1} = \sum_{m = 0}^{\infty} T^m, 
\end{equation}
which holds for $\|T\|<1$ in a suitable norm, and obtain
\begin{equation}
\hat{\rho}(t+\Delta t) \simeq \Big(1 + \mathcal{L}(t) \Delta t/2\Big)\sum_{m = 0}^{M} \Big(\mathcal{L}(t) \Delta t/2 \Big)^m \hat{\rho}(t), 
\label{eq:canck_vector}
\end{equation}
where we have introduced an upper cut-off in the sum. This expression allows us to propagate the density matrix forward in time. Importantly, we do not have to implement a matrix representation of the Liouvillian. We only need to implement its action on a density matrix, $\hat{\rho}\rightarrow \mathcal{L}\hat{\rho}$, and we only need to store the density matrix. Equation~(\ref{eq:canck_vector}) becomes particularly simple for $M=3$ and $M=7$, where it can be written on the product form
\begin{equation}
\hat{\rho}(t+\Delta t) \simeq \Big(1 + \mathcal{L}(t) \Delta t/2\Big)^2\Big(1 + (\mathcal{L}(t) \Delta t/2)^2\Big) \hat{\rho}(t), 
\label{eq:canck_vector_M=3}
\end{equation}
and
\begin{equation}
\begin{split} 
\hat{\rho}(t+\Delta t) \simeq  \Big(1 + &\mathcal{L}(t) \Delta t/2\Big)^2\Big(1 + (\mathcal{L}(t) 
\Delta t/2)^2\Big)\\
&\times\Big(1 + (\mathcal{L}(t) \Delta t/2)^4\Big) \hat{\rho}(t). 
\end{split}
\label{eq:canck_vector_M=7}
\end{equation}
Using this approach, we have time-evolved systems with up to 10 qubits with moderate numerical effort.

\begin{thebibliography}{42}
\bibitem [{aNo()}]{aNote2}%
	\BibitemOpen
	\href@noop {} {\bibinfo {title} {{For some oscillators, it may be difficult
				to modulate the total energy in time. For example, for a mechanical
				oscillator, we have $\hat H(t)=w(t)\left[\hat p^2/2m+k\hat
				x^2/2\right]=\hbar\omega_0(t)\left[\hat a^\dagger \hat a+ 1/2\right]$ for
				$\omega_0(t)=\sqrt{k(t)/m(t)}=w(t)\sqrt{k/m}$ with $m(t)=m/w(t)$ and
				$k(t)=w(t)k$, so both the mass and the spring constant would have to
				change.}}}\BibitemShut {Stop}%
\end{thebibliography}

%

\end{document}